\documentclass[10pt,letterpaper,twocolumn]{article} 

\usepackage{ol2}

\usepackage{amssymb,amsmath}
\usepackage{amsthm}
\usepackage{epsfig}
\usepackage{subeqnarray}
\usepackage{graphicx}
\usepackage{upgreek}
\usepackage{balance} 

\usepackage[numbers]{natbib}
\usepackage{notoccite} 
\usepackage{hyperref}  

\newcommand{\eq}[1]{{eq.(\ref{#1})}}
\renewcommand{\theequation}{{\arabic{section}.\arabic{equation}}}  

\begin{document}
\twocolumn[ 
\title{The Complete Positivity of Classical Polarization Maps}

\author{Omar Gamel}\email{ogamel@physics.utoronto.ca}\author{Daniel F. V. James} 
\address{Department of Physics, University of Toronto, 60 St. George St., Toronto, Ontario, Canada M5S 1A7.}

\begin{abstract}
Mueller and Jones matrices have been thoroughly studied as mathematical tools to describe the manipulation of the polarization state of classical light. In particular, the most general physical transformation on the polarization state has been represented as an ensemble of Jones matrices, as $\sum_i V_i \Phi V_i^\dag$. But this has generally been directly assumed with out proof by most authors. In this paper, we derive this expression from simple physical principles and the matrix theory of positive maps. 
\end{abstract}



]

\renewcommand{\theequation}{\arabic{equation}}
\newtheorem*{proposition*}{Proposition} 

\section{Introduction}
The polarization state of a classical beam of light can be mathematically represented in a variety of ways. The first description dates back over a century and a half to the work of Stokes \cite{stokes}, who introduced the four parameters which bear his name to specify the beam's polarization state \cite{brosseau98}. The Stokes parameters $S_\mu$, represent, in vector form, the power of the beam in various polarization modes. In the simple case of a fully polarized monochromatic deterministic beam, a Jones vector is commonly used to represent the state \cite{jones}.  

Further, powerful models that treat the electromagnetic field stochastically have been developed. Formalisms by Wiener \cite{wiener} and Wolf \cite{wolf54,wolf07,wolf03} allowed for probabilistic electric fields and the treatment of statistical optics via the polarization matrix, $\Phi$, formerly known as the coherency matrix \cite{oneill}. The Jones vector, or rather the projection operator formed by taking the outer product of the Jones vector with itself, can be seen as a special case of the polarization matrix. Therefore, we have two rival mathematical objects that describe classical polarization of a light beam, the Stokes vector $S_\mu$, and the polarization matrix, $\Phi$.
 
When an electromagnetic beam passes through an optical system, its state of polarization will, in general, be transformed. If one uses the Stokes vector to describe the state, then the transformation is represented via a Mueller matrix \cite{mueller}. If one represents the state through the polarization matrix, then transformation is represented via a Jones matrix \cite{jones}. 

It is generally assumed on physical grounds that the most general transformation on the state, in the polarization matrix formalism, can be represented as an ensemble of Jones matrix transformations \cite{kim87}, and based on this, the properties of the most general Mueller matrix can be derived \cite{simon82, gil00, luna} for the Stokes formalism. The proof of this assumption based on rigorous mathematical properties of the two formalisms has been lacking. Recently, Simon et al. \cite{simon10b} have addressed this problem based on properties of matrices derived from the Mueller matrices. 

In this paper, we address this problem differently. In section \ref{casignal} we review the concept of a complex analytic signal, the properties of which will be integral to our argument. We then review the polarization formalisms above in some detail in sections \ref{classicpolarization} and \ref{filters}. In section \ref{plmap} we show that provided only linear optical effects are allowed, then the most general transformation is indeed an ensemble of Jones matrix transformations. 

We base our main argument on basic properties of positive maps in two dimensions \cite{stormer, choi80} and simple physical assumptions about the state. The theorem on positive maps we use is similar to Choi's theorem for completely positive maps \cite{choi75} which is popular in quantum information theory \cite{nielsenchuang}. 

If one however allows nonlinear optical effects, particularly phase conjugation, then a more general transform is needed. We show the form of this alternative transformation. 

\section{Complex Analytic Signals}\label{casignal}

\subsection{Definition and Properties}

We begin by reviewing the concept of a complex analytic signal, which is fundamental to our argument. In what follows, our main reference is Mandel and Wolf, sec. 3.1 \cite{mandelwolf,wolf07}. Suppose we have a real-valued signal $x(t)$, which can be expressed via the Fourier synthesis integral: 
\begin{equation}
x(t) = \int^{\infty}_{-\infty} \tilde{x}(\omega) e^{-i\omega t}d\omega.
\end{equation}
Since the signal $x(t)$ is real, the Fourier spectrum $\tilde{x}(\omega)$ must satisfy $\tilde{x}(-\omega) = \tilde{x}^*(\omega)$, where the star denotes the complex conjugate. 

We note that the negative frequency components of the spectrum are fully determined by the positive frequency components. Therefore one can discard the former without loss of information. Thus, we define the complex analytic signal $z(t)$ as the signal synthesized only from the positive frequency components of $\tilde{x}(\omega)$. That is
\begin{equation}
z(t) \equiv \int^{\infty}_{0} \tilde{x}(\omega) e^{-i\omega t}d\omega.
\end{equation}

Alternatively, one may write the full Fourier synthesis equation for $z(t)$ as
\begin{equation}
z(t) \equiv \int^{\infty}_{-\infty} \tilde{z}(\omega) e^{-i\omega t}d\omega,
\end{equation}
where $\tilde{z}(\omega) \equiv \theta(\omega)\tilde{x}(\omega)$, and $\theta(\omega)$ is the Heaviside step function. From the above, it can be shown that
\begin{equation}
x(t) = 2\Re{[z(t)]},
\end{equation}
where $\Re$ denotes the real part (and $\Im$ the imaginary part) of a complex value.
For example, for the simple monochromatic signal $x(t)=cos(\Omega t)$, we have $\tilde{x}(\omega)=\frac{1}{2}[\delta(\omega + \Omega) + \delta(\omega - \Omega)]$, $\tilde{z}(\omega)=\frac{1}{2}\delta(\omega - \Omega)$, and $z(t)=\frac{1}{2}e^{-i\Omega t}$. 

We define $y(t)$ as twice the imaginary part of $z(t)$, that is
\begin{equation}
y(t) \equiv 2\Im{[z(t)]}.
\end{equation}
Then we can write the complex analytic signal as
\begin{equation}
z(t) = \frac{1}{2}[x(t) + i y(t)].
\end{equation}


One can show that the analytic property of $z(t)$ implies its real and imaginary parts above together form a \emph{Hilbert transform pair} \cite{titchmarsh}:
\begin{align}
y(t) &= \mathcal{H}[x(t)]\equiv \frac{1}{\pi}P \int^{\infty}_{-\infty} \frac{x(t')}{t-t'}dt',\nonumber\\
x(t) &= -\mathcal{H}[y(t)]= -\frac{1}{\pi}P \int^{\infty}_{-\infty} \frac{y(t')}{t-t'}dt',
\label{hilbertpair1}
\end{align}
where $P$ denotes the Cauchy principal value, and $\mathcal{H}$ is the Hilbert transform defined above. Note that the negative of a Hilbert transform is also its inverse transform. Considered as a function of complex $t$, the analytic signal $z(t)$ is analytic in the lower half of the complex $t$ plane. 

The relations in \eq{hilbertpair1} are identical, up to a sign, to the Kramers-Kronig relations \cite{kramers,kronig}, which relate the real and imaginary parts of the Fourier transform of a causal response function \cite{toll}. The difference between the case of a complex analytic signal and a causal response function is that the time and frequency domains have switched roles. In the case of the complex analytic signal, the frequency domain vanishes for negative arguments, and the time domain obeys the relation in \eq{hilbertpair1}, whereas for a causal response function, the time domain vanishes for negative arguments (equivalent to causality) and the frequency domain  obeys the Kramers-Kronig relations.

\subsection{Non-linearity of Complex Conjugation}\label{nonlinearconj}

Complex analytic signals make dealing with signals more convenient and streamline the mathematics, for the same reasons one prefers complex exponentials to trigonometric functions. For example, modulation, phase relationships and derivative properties become easier to deal with. Even though the real part of the signal is what sets the electric field, the complex part is not simply a convenience, it plays an important role in determining relative phases, and cross-correlation functions between multiple signals. Changing the imaginary component of the signal has observable effects.

To illustrate this, suppose we have two complex analytic signals given by $z_j(t) = \frac{1}{2}[x_j(t)+iy_j(t)]$, (i=1,2), that represent two wide-sense stationary stochastic processes with zero mean. The cross-correlation function between them is defined as
\begin{equation}
\Gamma_{12}(\uptau) \equiv \langle z_1^*(t)z_2(t+\uptau)\rangle,
\end{equation}
where $\langle \cdot \rangle$ denotes the ensemble average over different possible realizations.

One can show that $\Gamma_{12}(\tau)$ is itself a complex analytic signal. Note that since the generic complex analytic signal $z_1(t)$ contains only positive frequencies, its conjugate $z_1^*(t)$ only contains \emph{negative} frequency components. 

Suppose we have a conjugation device that transforms the stochastic process $z_1(t)$ to its complex conjugate, that is
\begin{align}
z_1(t) \rightarrow \underline{z}_1(t) &\equiv z_1^*(t) \nonumber\\
&=x_1(t) - iy_1(t).
\end{align}
We use notation $\underline{z}_1(t)$ for the transformed conjugate function to emphasize that it is just another complex function for which we can define correlations and a conjugate. We then define a secondary cross-correlation function:
\begin{align}
\Gamma_{\underline{1}2}(\uptau) &\equiv \langle \underline{z}_1^*(t)z_2(t+\uptau)\rangle, \\
&=\langle z_1(t)z_2(t+\uptau)\rangle.
\end{align}

While the cross-correlation function $\Gamma_{\underline{1}2}(\uptau)$ seems innocuous, using the generalized Wiener-Khintchine theorem \cite{mandelwolf,wiener,khintchine}, it can actually be shown to be identically zero. The technical reason for this being that there is no overlap between the spectra of $\underline{z}_1(t)$ and $z_2(t)$. That is, they do not share any nonzero frequencies in their Fourier spectrum; the former only has negative frequencies and the latter only has positive frequencies. 

The same conjugation device will then transform the cross-correlation function as
\begin{equation}
\Gamma_{12}(\tau) \rightarrow \Gamma_{\underline{1}2}(\uptau) = 0.
\end{equation}

So our hypothetical conjugation device would cause the cross-correlation function to always vanish. Given that the cross-correlation function is linear in $z_1(t)$, it seems the conjugation device cannot be linear in the physical sense. If it were linear, it would only send the zero functions to zero, or else it would be a trivial device that sends all functions to zero. Since neither is the case, we must conclude it is not a linear device. 

Indeed, known experimental techniques that conjugate phase, such as phase conjugate mirrors \cite{feinberg, boyd} are manifestly nonlinear in nature. So if we are restricting ourselves to linear devices, then the conjugation operation will not be allowed. 

Also note that if the operation $\mathcal{L}$ acting on electric fields is linear, then by definition it must satisfy
\begin{equation}
\mathcal{L}[\alpha z_1(t) + \beta z_2(t)] = \alpha\mathcal{L}[z_1(t)] + \beta \mathcal{L}[z_2(t)],
\label{linearEfield}
\end{equation}
where $\alpha$ and $\beta$ are arbitrary coefficients. However, the conjugation operation transforms the argument $\alpha z_1(t) + \beta z_2(t)$ to
\begin{equation}
[\alpha z_1(t) + \beta z_2(t)]^* = \alpha^* z_1^*(t) + \beta^* z_2^*(t).
\label{conj1}
\end{equation}
Comparing the last two equations, we see that the conjugation operation is not linear (in the sense of \eq{linearEfield}). It only satisfies the linearity property if we restrict ourselves to real coefficients $\alpha$ and $\beta$. For example, doubling the input will double the output, but if we have complex coefficients, we see that the complex conjugation operation is not linear.

Alternatively, one may argue the real and imaginary parts of a signal must form a Hilbert transform pair for it to be a complex analytic signal. By assumption, $z_1(t)$ is a complex analytic signal, and so
\begin{align}
y_1(t) &= \mathcal{H}[x_1(t)],\nonumber\\
x_1(t) &= -\mathcal{H}[y_1(t)].
\label{hilbertpair2}
\end{align}
Applying the conjugation operation to $z_1(t)$ means flipping the sign of $y_1(t)$. However, the conjugated $\underline{z}_1(t)$ does not describe a complex analytic signal because its imaginary part is not the Hilbert transform of its real part (there is a sign disparity due to the conjugation), and therefore is not admissible. In other words, complex conjugation in this context is an unphysical operation, inadmissible by the underlying formalism of classical stochastic linear optics. 

\section{Classical Polarization States}\label{classicpolarization}
Consider a classical beam of light propagating in the $z$ direction. The complex electric field values in the $x$ and $y$ direction are taken to be probabilistic ensembles given by complex analytic signals $E_1(\bold{r},t)$ and $E_2(\bold{r},t)$ respectively, where $\bold{r}$ is the position vector.

The polarization state of the beam of light is given by the 2$\times$2 polarization matrix $\Phi(\bold{r},t)$, defined as
\begin{equation}
\Phi_{ij}=\langle E_iE_j^* \rangle, \hspace{20pt} i=1,2.
\label{phidef}
\end{equation} 
where position and time dependence have been suppressed. If one thinks of $E_1$ and $E_2$ as random variables, then $\Phi$ is their variance-covariance matrix. 

Alternatively, the four element Stokes vector $S$ can be used to represent the polarization state \cite{stokes}. It is related to $\Phi$ by
\begin{align}
S_\mu &=Tr[\Phi \sigma^\mu ]=\Phi_{ij} \sigma^\mu_{ji}, \label{Seq}\\
\Phi &= \frac{1}{2}S_\mu\sigma^\mu, \label{Phieq}
\end{align} 
where $\sigma^0$ is the identity matrix, and $\sigma^1$, $\sigma^2$, and $\sigma^3$ are the three Pauli matrices $\sigma^z$, $\sigma^x$, and $\sigma^y$ respectively. Einstein summation notation has been used, i.e. repeated indices are summed over. Lowercase Latin letters run from 1 to 2 (corresponding to the two Cartesian components of the transverse field), while lowercase Greek letters run from 0 to 3. The polarization matrix or Stokes vector contain all the physical information about the polarization state of the beam \cite{brosseau98}, and are different ways of mathematically representing the same information.

\section{Filters}\label{filters}
\subsection{Simple Filters}

When the beam interacts with a linear optical element, generically called a filter, its polarization state is transformed. The most basic type of filter, which we call a simple filter, linearly transforms the transverse electric field vector via the well-known Jones matrix \cite{jones}, denoted $T$, through simple matrix multiplication:
\begin{equation}
E_i'=T_{ij}E_j.
\end{equation}
This is equivalent to the following unitary transformation on the polarization matrix:
\begin{equation}
\Phi' = T\Phi T^\dag.
\label{unitary}
\end{equation}
Turning to the Stokes vector, any linear transformation on the state must have the form
\begin{equation}
S_\mu'=M_{\mu\nu}S_\nu,
\label{mueller}
\end{equation}
where $M_{\mu\nu}$ is the 4$\times$4 Mueller matrix \cite{mueller}. Equations (\ref{Seq}), (\ref{Phieq}) and (\ref{unitary}) then imply that the Mueller matrix corresponding to a simple filter can be expressed as \cite{brosseau98, kim87, gil00}
\begin{align}
M_{\mu\nu} &= \frac{1}{2}Tr[\sigma^\mu T\sigma^\nu T^\dag] \label{simpleMueller1} \\
 &= A(T\otimes T^*)A^{-1} \nonumber\\
&= \frac{1}{2}A(T\otimes T^*)A^\dag,
\label{simpleMueller2}
\end{align}
%
%
%
where $\otimes$ is the tensor product, and A is the matrix whose rows are the vectorization of the Identity and Pauli matrices, given by
\begin{equation}
A=\left[ \begin{array}{cccc} 1 & 0 & 0 & 1 \\ 1 & 0 & 0 & -1 \\ 0 & 1 & 1 & 0 \\ 0 & i & -i & 0 \end{array} \right].
\end{equation} 
Mueller matrices of this kind are called \emph{pure Mueller} matrices, or \emph{Mueller-Jones} matrices \cite{gil00}. 

Natural questions that arise at this point are how does one represent the most general type of optical filter and what are the relevant Jones and Mueller matrices?

\subsection{General Filters}

It has been a generally held axiom, motivated by physical intuition, that the most general kind of optical filter is an ensemble of simple filters \cite{kim87, simon82, gil00}. Therefore, the transformation on the polarization matrix is then represented by an ensemble of Jones matrices, rather than a single one, through the expression
\begin{equation} 
\Phi' = \sum_e p_eT_e\Phi T_e^\dag,
\label{comppos}
\end{equation}
where $e$ is the index over the elements of the ensemble, and $p_e$ is the probability of realizing the $e^{th}$ ensemble element ($\sum_e p_e = 1$).

From \eq{simpleMueller1} we find that the Mueller matrix for a general filter is given by
\begin{align}
M_{\mu\nu} &= \sum_e p_e \frac{1}{2}Tr[\sigma^\mu T_e\sigma^\nu T_e^\dag] \\ 
&=\sum_e p_e M_{\mu\nu}^{(e)},
\label{generalMueller}
\end{align}
where the $M_{\mu\nu}^{(e)}$ are pure Mueller matrices, each derived from a single Jones matrix in the ensemble. This expression tells us that any physically admissible Mueller matrix is given by a convex linear combination of pure Mueller matrices. That is, the set of all Mueller matrices is the convex hull \cite{chull} of pure Mueller matrices. 

At first glance, \eq{mueller} seems to suggest that all matrices $M_{\mu\nu}$ that map the set of physical Stokes vectors into itself are physical Mueller matrices. This condition of mapping Stokes vectors to Stokes vectors turns out to be a necessary but \emph{not sufficient} condition for a physical Mueller matrix \cite{simon82, gil00,simon10}. Another necessary condition has to do with the positivity of the polarization matrix. References \cite{simon10, simon10b} show this through the use of matrices derived by rearranging entries of the Mueller matrix, a basis of 16 unitary matrices, and the linear independence of components of the electric field vector. They also make use of the beam-coherence-polarization matrix \cite{gori98}, which measures coherence between two different spatial points.

In the next section, we derive \eq{comppos} in a much simpler manner from basic mathematical properties, making use of an interesting theorem on positive linear maps in C*-algebras \cite{stormer, choi80}. 
\section{Positive Linear Map Approach}\label{plmap}
\subsection{Axioms}
Let $\mathcal{F}(\Phi)$ denote the operation of a general linear filter on the polarization matrix $\Phi$. Our goal is to find the form of $\mathcal{F}(\Phi)$ based on a few intuitive axioms. We assume $\mathcal{F}(\Phi)$ satisfies the following simple axioms:

\begin{enumerate}
\item
It is linear in $\Phi$. That is, for any set of coefficients $\{ c_i \} $, we have $\mathcal{F}\Big(\sum_i c_i \Phi_i \Big) = \sum_i c_i \mathcal{F}(\Phi_i).$
%
\item
It is positive. That is, $\Phi' = \mathcal{F}(\Phi)$ is a positive 2$\times$2 matrix for any positive 2$\times$2 matrix $\Phi$.

\item
It is composed of linear operations in the electric field. That is, $\mathcal{F}$ must only use operations that satisfy \eq{linearEfield}.
%
\end{enumerate}

The validity of the first axiom is based on the superposition principle, assuming of course the different polarization matrices $\Phi_i$ correspond to independent electric fields (i.e. the electric fields underlying the polarization matrices being added are mutually uncorrelated) \cite{bornwolf}. The second axiom holds true since the output of $\mathcal{F}$ is a physical polarization matrix, and therefore must be positive. The third axiom relies on the fact that we are only allowing linear optical devices, which ultimately must act in a linear fashion on the electric field.

We can then make the following proposition:
\begin{proposition*}
The simplest expression for $\mathcal{F}(\Phi)$ does not include the conjugate of the input matrix, $\Phi^*$. In other words, $\mathcal{F}$ is holomorphic.
\end{proposition*}

To justify this proposition, we note from the definition of $\Phi$ in \eq{phidef} that applying the complex conjugation operator to $\Phi$ is equivalent to applying it to the electric fields. That is, it is equivalent to $E_j \rightarrow E_j^*, j=1,2$. The electric field of a beam \emph{at a given point} must be given by the real component of the complex analytic electric field, expressed as
\begin{equation}
E_j(t) = \frac{1}{2}[E_j^{(r)}(t) + iE_j^{(i)}(t)], \hspace{20pt} j=1,2.
\end{equation}

Based on our earlier discussion on the nonlinearity of the phase conjugation operation when applied to complex analytic signals in section \ref{nonlinearconj}, we see that indeed the conjugation operation in the expression $\Phi^*$ would be nonlinear. Using the third axiom of linearity in the electric field, we see that our proposition is justified. 

\subsection{Derivation}

To derive the form of $\mathcal{F}$, we take the first two axioms. Together, they state that $\mathcal{F}$ is a positive linear map that maps the space of $2\times2$ matrices onto itself. In the theory of C* algebras, the most general form for such a map is well known \cite{stormer, choi80}. It is given by
\begin{equation}
\mathcal{F}(\Phi) = \sum_i V_i \Phi V_i^\dag + \sum_j W_j \Phi^t W_j^\dag ,
\label{osrtranspose}
\end{equation} 
where $\Phi^t$ is the transpose of $\Phi$, and $V_i$, $W_j$ are arbitrary matrix operators of suitable dimension. Recall that $\Phi$ is Hermitian as can be seen from \eq{phidef}. Therefore its transpose is equal to its complex conjugate ($\Phi^t=\Phi^*$). One must mention here that \eq{osrtranspose} is for a \emph{positive} map, not the more commonly used \emph{completely positive} map. There is a subtle difference between the two; a positive map transforms a positive matrix to a positive matrix, whereas a completely positive map has the additional requirement that it be positive even when tensored with an identity operation of any dimension \cite{choi75}. 

The transpose is the most well known example of an operation that is positive, but not completely positive \cite{nielsenchuang}, hence there is no surprise it appears in \eq{osrtranspose}. Note that \eq{osrtranspose} only holds for dimensionality $2\times2$. No analogous expression is known for higher dimensional positive maps \cite{stormer, choiconversation}.

Applying our proposition, which is based on the third axiom, to \eq{osrtranspose}, we see that the term involving the transpose must vanish. This implies that the $W_j$ can no longer be arbitrary, and we must set $W_j = 0 \ \forall j$. Then we have the final form of the most general physical transformation upon the polarization matrix $\Phi$ as
\begin{equation}
\mathcal{F}(\Phi) = \sum_i V_i \Phi V_i^\dag.
\label{osrfinal}
\end{equation}
Interestingly, the expression in \eq{osrfinal} describes the general form of a completely positive map. However, we derived it in this particular circumstance without making use of Choi's well known theorem on completely positive maps \cite{choi75}, but through constraints of physical linearity applied to the less restrictive positive map. Dropping the transpose term in the process, we are left with a completely positive map after all. Note that in the field of quantum information, quantum channels are completely positive transformations that act on the density matrix \cite{nielsenchuang}, where the operators $V_i$ are the Kraus operators \cite{kraus}.  

A potential point of confusion must be clarified here. We mentioned that we discarded the conjugation operation (and by extension the transpose) because it is not linear. Yet \eq{osrtranspose} is the expression for a positive \emph{linear} transformation. The confusion is resolved when we note that the conjugation operation was discarded because it is not linear \emph{in the electric field} $E$, whereas \eq{osrtranspose} is linear \emph{in the polarization matrix} $\Phi$ (and not the electric field). So it is linearity in two different senses.

To recap, we have shown that \eq{osrfinal} describes the most general operation upon a polarization matrix $\Phi$ in the context of linear optics. Comparing \eq{osrfinal} with \eq{comppos} and setting $V_i = \sqrt{p_i}T_i$, we see that, as expected, $\mathcal{F}(\Phi)$ is of the form of a Jones matrix ensemble. 

Alternatively, if one just admits conjugation operations through a \emph{specific} nonlinear apparatus, then \eq{osrtranspose} is the most general operation on the polarization matrix $\Phi$. If we allow \emph{any} nonlinear operation, then even the first axiom no longer holds, and more general expressions must be used.

\section{Conclusion}

We have proven the the validity of the expression in \eq{comppos} as the most general physical transformation on a polarization matrix, using some basic mathematics and simple assumptions. This puts the assumption that an ensemble of Jones matrices is the most general linear optical filter on more solid ground, and illustrates exactly where the assumption will break down if we relax the linearity requirement.

We have also given physical reasons why the transpose map is inadmissible, despite its preservation of positivity, equating it to the unphysical (and nonlinear) conjugate map. This treatment will break down in higher dimensions, since \eq{osrtranspose} only applies in the case of $2 \times 2$ matrices. Moreover, if one admits conjugation operations through nonlinear optical devices, we find the more general \eq{osrtranspose} is the most general transformation.

\section*{Acknowledgements}
We thank Professor Man-Duen Choi at the University of Toronto Department of Mathematics for his help. This work was funded by the Natural Sciences and Engineering Research Council of Canada (NSERC), and the Ontario Ministry of Training, Colleges, and Universities.

\bibliographystyle{unsrt}
\balance 
{\footnotesize
\bibliography{CPMapsBib}
}

\end{document}